\documentclass{PoS}

\newcommand{\be}{\begin{equation}}
\newcommand{\en}{\end{equation}}
\newcommand{\bi}{\begin{itemize}}
\newcommand{\ei}{\end{itemize}}
\newcommand{\bea}{\begin{eqnarray}}
\newcommand{\ena}{\end{eqnarray}}

\newcommand{\hbo}{\hbox to 1 true cm {\hfill } }
\newcommand{\tr}{\hbox{tr}}

\title{A fresh look at the confinement mechanism}

\ShortTitle{confinement versus cooling }

\author{\speaker{Kurt Langfeld} \\ % \thanks{A footnote may follow.}\\
        School of Computing \& Mathematics, University of Plymouth \\
        Plymouth PL4 8AA, United Kingdom \\
        E-mail: \email{kurt.langfeld@plymouth.ac.uk}}

\abstract{Topological configurations, monopoles and vortices, successfully 
describe quark confinement and the spontaneous breakdown of chiral symmetry. 
Despite their infinite action, these configurations are relevant due to 
a subtle cancellation between action and entropy. A natural explanation 
for this intrinsic fine-tuning is that smooth low action configurations  
exist which confine and which appear as singular topological objects 
in certain gauges. To reveal these confining semi-classical configurations, 
a new cooling method is proposed which largely reduces the action while 
preserving the asymptotic quark-antiquark potential. First numerical 
results for a SU(2) gauge theory show that confining configurations
with an average plaquette as high as $0.95$ do exist. 
}

\FullConference{International Workshop on QCD Green's Functions, Confinement, and Phenomenology - QCD-TNT09\\
		 September 07 - 11 2009\\
		 ECT Trento, Italy}

\begin{document}

\section{Introduction}
Two low energy phenomena of Quantumchromodynamics (QCD)
predominantly dictate the properties of matter under normal conditions:
quark confinement and the spontaneous breakdown of chiral symmetry.
It was already realised in the mid seventies that Yang-Mills theories
admit topologically non-trivial configurations with finite
action called 
instantons~\cite{Belavin:1975fg,Callan:1977gz,tHooft:1976fv,Shuryak:1981ff}.
It soon became clear that a dilute gas~\cite{Callan:1977gz}
or a liquid~\cite{Shuryak:1981ff,Diakonov:1983hh} of instantons
may explain the spontaneous breakdown of chiral symmetry.
After the advent of computer studies of Yang-Mills theories by means
of lattice gauge theories, configurations of low action have been
constructed by cooling. It has been reported in that straightforward
cooling of lattice configurations annihilates the string tension
and therefore quark confinement~\cite{Polikarpov:1987yr}. Direct
evidence of instanton generated confinement has not yet been
established.

The wide spread belief is that a different type of configurations which 
populate the Yang-Mills vacuum is responsible for quark confinement. 
For example, colour magnetic monopoles
generically exist in certain gauges and might generate quark
confinement be means of the dual Meissner
effect~\cite{'tHooft:1981ht,Kronfeld:1987ri}. Alternatively,
a vacuum populated with percolating vortices
which carry flux in the centre of the non-Abelian gauge group
would equally well explain quark
confinement\cite{Cornwall:1979hz,Mack:1980zr}. Equipped with rapidly
increasing computational resources, these ideas were put to the test
using lattice gauge simulations. It turned out that the early definitions
of the centre vortices failed the quantitative test: the properties of those
vortices were strongly dependent on the UV cutoff of the Yang-Mills theory. 
An important novel definition of the vortex matter was firstly proposed
in~\cite{DelDebbio:1996xm,Del Debbio:1996mh,Del Debbio:1998uu}, and the
connection of those vortices has been established using lattice gauge
simulations. It was  pointed out for the first time
in~\cite{Langfeld:1997jx,Engelhardt:1998wu}
that these vortices are sensible in the continuum limit when the
UV regulator of the theory is removed. In particular for the gauge group
SU(2), this vortex texture of the vacuum explains many of the low energy
features of Yang-Mills theory. In particular,  deconfinement
at finite temperatures appears as a vortex depercolation
transition~\cite{Langfeld:1998cz,Engelhardt:1999fd,Langfeld:2003zi}, and 
a close connection between quark confinement and spontaneous chiral symmetry breaking has been found: Removing vortices restores
chiral symmetry as realise from the quark
condensate~\cite{de Forcrand:1999ms},  from the spectral density of
the quark operator~\cite{Gattnar:2004gx} or from an inspection
of the Landau gauge quark propagator~\cite{Bowman:2008qd}.
If the discretised version of the quark operator possesses acceptable
chiral properties even for singular vortex background fields,
vortex only configurations also show spontaneous chiral symmetry
breaking~\cite{Hollwieser:2008tq}. Up to now, the vortex picture
for the gauge group SU(3) is less compelling: Centre projection
falls short to reproduce the full string tension ('$2/3$'-problem)
~\cite{Langfeld:2003ev},
and the tight connection between confinement and spontaneous chiral
symmetry breaking seems to be lifted~\cite{Leinweber:2006zq}. The
'$2/3$'-problem is absent if the Laplacian Centre gauge is used 
to define the vortex matter~\cite{deForcrand:2000pg}. As pointed out in~\cite{Langfeld:2001nz,Langfeld:2003ev}, however, the
Laplacian vortex matter is not independent from the UV regulator.
The practical consequences are that the vortex removed model theory
has lost the property of asymptotic freedom and generically suffers
from a low signal to noise ratio. Further SU(3) studies are needed
to resolve these issues. 

Both, in SU(2) and SU(3), the scaling of vortex matter arising from the
maximal centre gauge is a remarkable feature
of the theory: the infinite action of the vortex world sheets is balanced by
an equally large entropy~\cite{Gubarev:2002ek,Zakharov:2003vh}. Since these  
vortices
are defined in a particular gauge, i.e., the so-called maximal centre gauge,
a natural explanation of this intrinsic fine tuning would be that there is 
a smooth low action analogue of the vortex
configurations which equally well induce confinement.
Unravelling these semi-classical configurations would facilitate
an analytic first principle understanding of confinement.

\vskip 0.3cm
In this paper, we report a novel cooling technique which strongly
reduces the action of the lattice configurations, but
which preserves the long distance behaviour of the static quark
anti-quark potential and thus quark confinement.

\section{Confinement preserving cooling}

Throughout this paper, we adopt the lattice regularisation
of SU(2) Yang-Mills theory. Degrees of freedom are the matrices
$U_\mu (x)$ associated with the link of the space-time lattice with
lattice spacing $a$. We use the Wilson action
\be
S_\mathrm{wil}
\; = \; - \beta  \sum_{x,\, \mu > \nu} \frac{1}{2} \, \tr \, \Bigl[
U_\mu (x) \, U_\nu (x+\mu) \, U^\dagger _\mu (x+\nu) \, U^\dagger _\nu (x)
\Bigr] \; .
\label{eq:1}
\en
For our cooling method below, we need to address the static quark-antiquark
potential $V(r)$ in terms of un-smeared Wilson loops. The expectation
value of a rectangular Wilson loop $W[r,t]$ with extent $r$ in spatial
and $t\gg 1/m_G$ in time direction is related to the static potential by:
\be
\Bigl\langle W[r,t] \Bigr\rangle \; = \; \Big\vert \langle 2 \vert
\Omega _2 \rangle \Big\vert ^2 (r) \; \exp \Bigl\{ - V(r) \, t \Bigr\} \; ,
\label{eq:2}
\en
where $m_G$ is the mass gap of the theory, $\vert 2 \rangle$ is the state
of two heavy, axial-dressed
quarks~\cite{Heinzl:2007kx,Heinzl:2008tv,Heinzl:2008bu} and
$\vert \Omega _2 \rangle $ is the true ground state in the quark-antiquark
channel. For axial-dressed quarks, the overlap $\vert \langle 2 \vert
\Omega _2 \rangle \vert^2 (r)$ is a small number which
exponentially decreases with increasing $r$~\cite{Heinzl:2008tv}. It is therefore strongly advisable to use
Coulomb dressing or dressing via APE smearing, which both improve on the
overlap, if the  aim is the calculation of the static potential. 

\vskip 0.3cm
The idea central to our novel cooling is to minimise the
action (\ref{eq:1}), but to keep fixed the static potential $V(r\ge r_c)$.
For our method, it will turn out below that fixing the potential
for $r=r_c$ is sufficient to keep $V(r \ge r_c)$ unchanged. Hence, 
$r_c$ is a new scale which comes into play, and which is reminiscent of the
so-called cooling radius employed in~\cite{de Forcrand:1997sq} for elaborate cooling. A straightforward but wrong idea would be
to constrain the Wilson loop expectation value 
$\langle W(r=r_c,t) \rangle $ during cooling. Since the size of the right hand side
of (\ref{eq:2}) is dominated by the overlap rather than the exponential,
this would merely imply that we freeze the poor overlap. This is not what we
want. By contrast, cooling the lattice configuration should
increase the overlap of the trial state with the true ground state.
To this aim, we consider the ratio
\be
\frac{ \langle W[r , t+1] \rangle }{ \langle W[r,t] \rangle }
\; = \; \exp \Bigl\{ - V(r) \, a \Bigr\} \; .
\label{eq:3}
\en
A valid method would be to constrain the above ratio for $r=r_c$.
Since the numerator and the denominator consist of an average over
many lattice configurations, this method is hardly feasible.
To circumvent this practical problem, we firstly note that the
Wilson loop $W[r,t](x)$ for a given lattice configuration depends
on the position $x$ on the lattice. The configuration average, however,
is position independent due to translation invariance~\footnote{In practical
calculations, we sum over all positions to increase the signal-to-noise
ratio.}. Consider now two Wilson loops at $x$ and $y$ with a separation
much bigger than the inverse mass gap. We then find
$$
\Bigl\langle W[r,t](x) \; W[r,t](y) \Bigr\rangle \; \approx \;
\Bigl\langle W[r,t](x) \Bigr\rangle \; \Bigl\langle
W[r,t](y) \Bigr\rangle , \hbo \vert x-y \vert \gg \frac{1}{m_G} \; .
$$
For pure Yang-Mills theory, the mass gap $m_G$ is set by the mass
of the lightest $O^{++}$ glueball implying that $\xi := 1/m_G$ is short ranged.
This implies that Wilson loops separated by distances larger than $\xi$
are statistically independent. For large lattices, i.e., $Na \gg \xi$,
we might therefore approximate the average over lattice configurations by an
average over space time (and orientation; not shown)
\be
\Bigl\langle W[r,t](x) \Bigr\rangle \; \approx \;
\frac{1}{6N^4} \sum _x W[r,t](x) \; .
\label{eq:4}
\en
The total action which we are going to minimise is given by
\be
S _\mathrm{all} \; = \; S _\mathrm{wil} \; + \; \lambda \;
\frac{ \sum _x  W[r_c,t+1](x) }{ \sum _x  W[r_c,t](x) } \; ,
\label{eq:5}
\en
where $\lambda $ is a Lagrange multiplicator which ensures that
\be
R:= \; = \; \frac{ \sum _x  W[r_c,t+1](x) }{ \sum _x  W[r_c,t](x) } \; = \;
\hbox{constant}
\label{eq:6}
\en
during cooling. Any sufficiently large value for $t$ should work. 
We have chosen 
$$ 
t \; = \; r_c 
$$
throughout this paper. 
The action (\ref{eq:5}) can be minimised by standard means such as
steepest decent method. The outline of the cooling approach is as follows:
\bi
\item[1] Generate a thermalised lattice configuration $\{ U_\mu \}$;
\item[2] Calculate the ratio $R$ in (\ref{eq:6});
\item[3] Calculate a change $\delta U$ of the link configuration
which reduces the total action (\ref{eq:5});
\item[4] Adjust the Lagrange multiplier $\lambda $ to ensure that $r$
has not changed when you implement the change $\delta U$;
\item[5] Repeat steps 2-5 until the action (\ref{eq:5})
does not change anymore;
\item[6] Start again with step 1 to generate an new cooled configuration.
\ei
In order to show that this approach is feasible,
we here just report the change of the action $\delta S$ when the particular
link $U_\mu (x)$ is changed to $\delta U$ while all other links
are kept fixed:
\bea
\delta S &=& \delta U_\mu (x) \; \Bigl[ - \beta \,
B_\mu (x) \; + \; \lambda \,
\frac{ C_\mu[r_c,r_c+1] (x) }{   \sum _x  W[r_c,r_c](x) }
\; - \; \lambda \,
\frac{ \sum _x  W[r_c,r_c+1] (x) }{   \{ \sum _x  W[r_c,r_c](x) \}^2  } \;
C_\mu[r_c,r_c] (x) \Bigr] ,
\nonumber \\
B_\mu (x) &=& \sum _{\nu \not=\mu} U_\nu (x+\mu) \, U^\dagger _\mu (x+\nu) \,
U^\dagger _\nu (x) \; .
\label{eq:8}
\ena
Thereby, the $B_\mu (x)$ are the well known {\it staples}, and
$C_\mu[r_c,t] (x)$ is the sum of cut-open Wilson loops $W[r_c,t]$ starting
at point $x+\mu $ and ending at point $x$.

\section{Numerical results:}

\begin{figure}
\vskip -4cm
\begin{center}
\includegraphics[width=9cm]{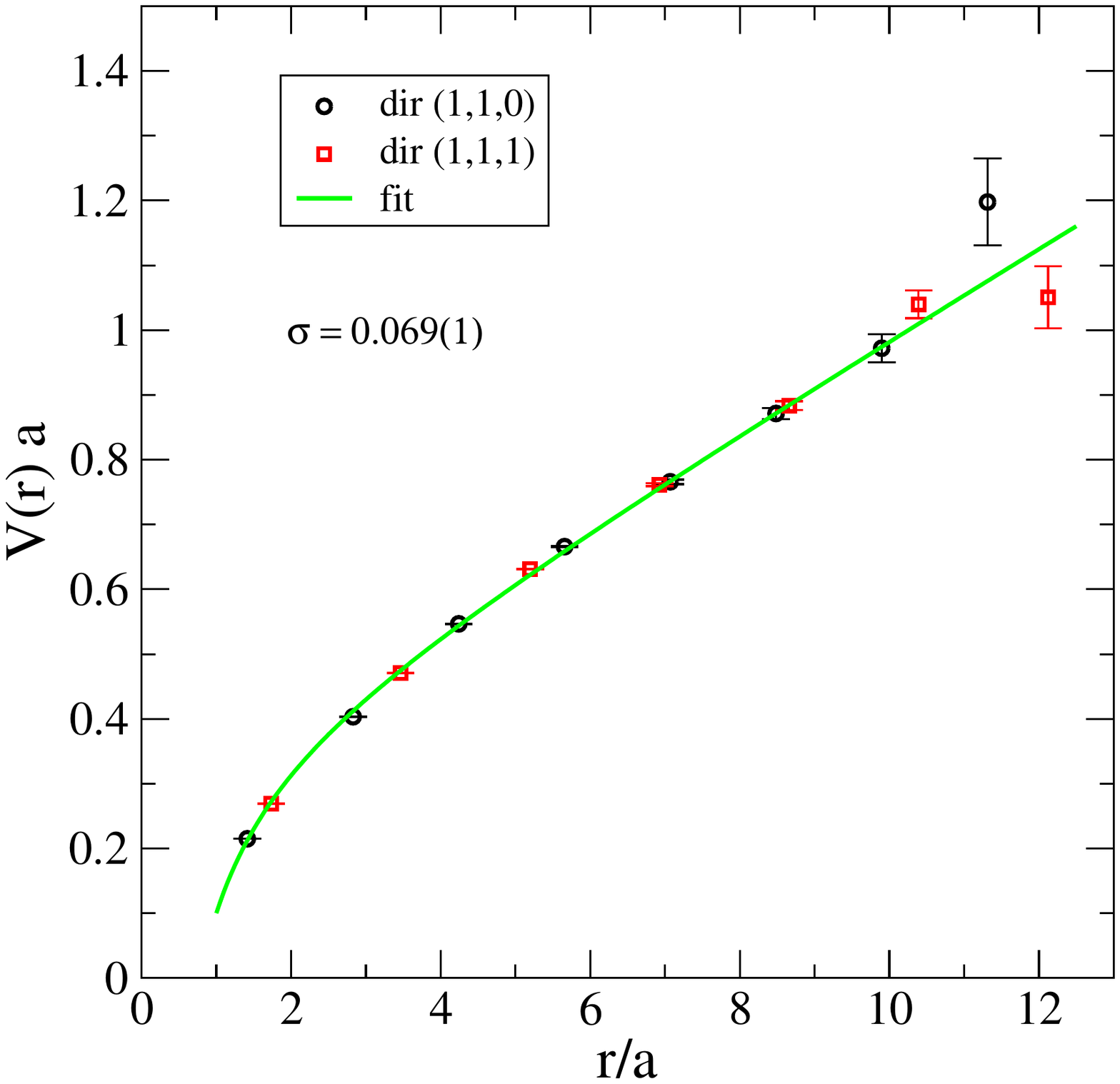}
\caption{ Static quark-antiquark potential obtained from cooled
configurations. }
\label{fig:1}
\end{center}
\end{figure}

First numerical results were obtained for a $16^4$ lattice for
a gauge group SU(2) with $\beta = 2.4$. The cooling radius was fixed to
$r_c=3$. We find that after $150$ cooling steps further changes
of the action are tiny.

\begin{figure}
\begin{center} 
\includegraphics[width=11cm]{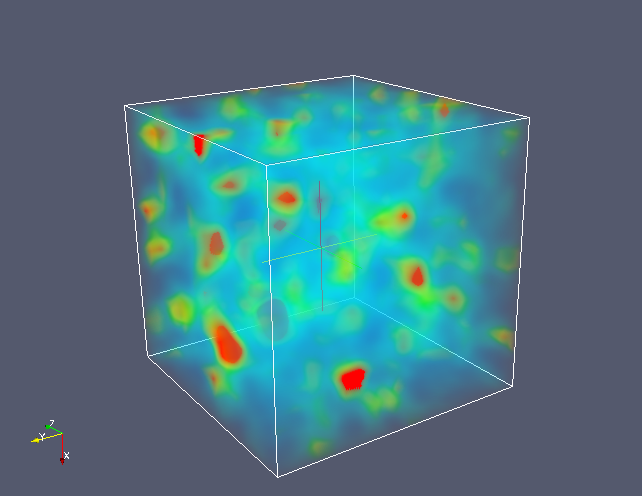}
\caption{ Action density obtained from cooled
configurations with cooling radius $r_c=3a$, $\beta =2.4$, $20^4$-lattice.  }
\label{fig:2}
\end{center}
\end{figure}

We firstly checked whether we recover the correct asymptotic
behaviour of the static quark-antiquark potential from cooled configurations.
To this aim, we evaluated the expectation value of rectangular
Wilson loops averaged over cooled configuration. The would-be static
potential $\bar{V}(r)$ is then obtained from
$$
- \ln \, \Bigl\langle W[r,t] \Bigr\rangle _\mathrm{cooled} \; = \;
\bar{V}(r) \, t \; - \; \hbox{function}(r) \; , \hbo t \gg 1/m_G .
$$
For large values $t>t_f$, this function is fitted to a linear function.
If the $\chi^2/\mathrm{dof}$ is not satisfactory, a possible cure would be
to increase $t_f$.
We studied the cases where the quark and the antiquark
are aligned along the main cubic axis ($(100)$-direction) as well as
along the diagonals $(110)$ and $(111)$. It turns out that the linear
fit was never satisfactory for the orientation $(100)$ of quark and
antiquark while this problem did not occur for the diagonal directions.
Further studies will be needed to trace out the origin of this effect.
For the moment, we keep in mind that the cooled configurations
might be plagued by sizable rotational symmetry breaking effects.
Because of the constraint during cooling, we ensure that
$\bar{V}(r_c) = V(r_c)$. The crucial question is whether
$\bar{V} = V$ also holds for $r>r_c$. Our numerical result obtained from
$640$ cooled configurations is shown in figure~\ref{fig:1}.
The fitted string tension is given by $0.069(1)$ which compares quite well 
with value $0.075(2)$ from full configurations. After a sufficiently large 
amount of cooling steps ($>150$ for my setup), the action density 
seems to stabilise. The constraint during cooling prevents the 
configurations to become the perturbative vacuum with trivial plaquettes 
only. An example of such a cooled configuration (cooling radius $r_c=3a$, $\beta =2.4$) is shown in figure~\ref{fig:2}. We clearly observe a lump type structure in the
vacuum. The overall achieved average plaquette of this configuration 
has been $\approx 0.95$.

\section{Conclusions} 
Over the last two decades, topological configurations, such as
monopoles and vortices, have been successfully linked to the confinement
phenomenon in Yang-Mills theories. It has been observed in the recent past
that those degrees of freedom also describe the spontaneous breakdown
of chiral symmetry. These configurations are relevant due to a subtle
cancellation between action and entropy. A natural explanation for this 
intrinsic fine tuning is that this singular configurations have a low 
action counter part in a different gauge. In order to unravel these 
semi-classical configurations, a new type of cooling was introduced 
which largely reduces the action but preserves the asymptotic behaviour 
of the quark potential. The viability of the new cooling method 
has been explored for the SU(2) gauge theory for lattices as big as 
$20^4$. A lump type vacuum structure has been observed using the 
action density. 

Future studies will address the questions: 
Does the action density of the cooled configurations scale like 
$a^4$ ($a$ the lattice spacing) implying a finite action density 
in the continuum limit? What is the topological charge of the observed 
lumps, and what is their relation to instantons and 
calorons~\cite{Kraan:1998sn}?
Does the observed structure match with monopoles and vortices 
in the corresponding gauges? 

\acknowledgments 
The numerical computations were carried out at the HPC-Centre of 
the University of Plymouth.

\end{document}